\documentclass{iopconfser}

\usepackage{graphicx}
\usepackage{float} %for [H] in figures
\usepackage{amsmath}
\usepackage{subcaption}

\usepackage[colorlinks=false]{hyperref}
\hypersetup{
	pdftitle    = Radiation Safety Challenges in Plasma Accelerators
	pdfauthor   = Simon Bohlen,
}
\usepackage[separate-uncertainty]{siunitx}

\usepackage{lmodern}

\usepackage{tabularx}
\usepackage{comment}

\begin{document}

\title{Radiation safety challenges in plasma accelerators}

%\author{S.~Bohlen$^{1}$ , A.~Leuschner$^{1}$, T.~Liang$^{1}$, M.~Thévenet$^{1}$, I.~Yeh$^{1}$}
\author{S.~Bohlen$^{1,*}$,
M.~Kirchen$^{1}$,
T.~Liang$^{1}$,
A.~Leuschner$^{1}$,
A.~R.~Maier$^{1}$,
A.~Martinez~de~la~Ossa$^{1}$,
E.~Panofski$^{1}$,
K.~Schubert$^{1}$,
M.~Thévenet$^{1}$,
P.~A.~Walker$^{1}$,
I-L.~Yeh$^{1}$ 
and S.~Zander$^{1}$}
%\author{S.~Bohlen$^{1}$, et al.}

\affil{$^{1}$Deutsches Elektronen-Synchrotron DESY, Notkestraße 85, 22607 Hamburg, Germany}
\affil{$^*$Author to whom any correspondence should be addressed.}

\email{simon.bohlen@desy.de}
\begin{abstract}
Plasma accelerators are rapidly evolving toward user-relevant machines with increasing repetition rates, particle energies and average beam powers. Despite their compact size, the operational characteristics of plasma accelerators are comparable to those of radio-frequency linacs, involving the continuous generation and dumping of electron bunches. However, beam properties and loss patterns can differ substantially from those of conventional accelerators, leading to radiation safety considerations dominated by high peak charges and distributed beam losses relevant for both personnel protection and machine integrity.
Using established scaling laws, we show that significant dose rates already occur at electron energies of several MeV, underscoring the relevance of radiation protection even for comparatively low-energy plasma accelerators. Based on a combination of Monte Carlo and particle-in-cell simulations, supported by radiation measurements from plasma accelerator experiments at DESY, we analyze typical radiation fields with a particular focus on radiation generated close to the plasma source. These findings highlight the need for dedicated shielding and beam-dump concepts tailored to plasma accelerators, especially in view of increasing average beam powers and future application-oriented operation.
\end{abstract}

\section{Introduction}
Plasma accelerators \cite{Tajima1979LaserAccelerator,Chen1985AccelerationPlasma} are currently at a pivotal stage, transitioning from proof-of-principle experiments toward first concepts for user-oriented machines and practical applications. This development is driven by steadily increasing average beam powers \cite{Rovige2020DemonstrationAccelerator,Maier2020DecodingAccelerator,Bohlen2022StabilityAccelerators}, enabled by advances in laser technology and supported by an improved understanding of the fundamental scalability offered by plasma-based acceleration schemes \cite{DArcy2022RecoveryAccelerator}. From a radiation safety perspective, this evolution necessitates a more comprehensive assessment of the radiation fields produced by such machines.

From an accelerator physics point of view, plasma-based devices operate as linear accelerators, characterized by the continuous generation and dumping of fresh electron bunches. This stands in contrast to synchrotrons and storage rings, where particle bunches circulate for extended periods (several hours) before being extracted or lost. As a consequence, the total charge generated, and thus the associated radiation dose from electron losses, can be orders of magnitude higher in linear accelerators than in synchrotron-based facilities. This difference is well illustrated by a comparison of two large user facilities at DESY: the PETRA~III synchrotron~\cite{Drube2016TheExtension} and the FLASH free-electron laser~\cite{Ackermann2007OperationWindow}. While beam losses at PETRA~III are on the order of $\SI{1e15}{electrons\,per\,year}$, at FLASH a comparable amount of charge is generated and subsequently dumped within only a few seconds of operation.

In plasma accelerators, this challenge is further amplified by broad energy spectra and angularly distributed electron emission, which complicate controlled beam transport and radiation confinement. Additional challenges arise from the beam properties typically observed in plasma-based acceleration schemes. Beam parameters are often less stable than in conventional radio-frequency linacs \cite{Emma2021FreeProspects}, owing to the fact that the plasma wakefield accelerating the particles is generated anew on each shot through partially non-linear processes. This results in less controlled operating conditions and more unpredictable beam-loss scenarios. Moreover, radiation generation is not limited to the accelerated high-energy electron bunch alone. Electrons within the plasma itself may be heated either by intense laser pulses or by charged particle beams, thereby contributing to the overall radiation field \cite{Kaganovich2008ObservationWakefield,Yang2017ThreeQuestion,Garrett2025ExcitationAcceleration}. In beam-driven plasma accelerators, the presence of a spent driver bunch with a large energy spread \cite{Pena2024EnergyAccelerator} further complicates controlled charge transport and efficient beam dumping.

In this paper, the main radiation safety challenges associated with plasma accelerators are discussed by combining established radiation physics concepts with experimental observations from plasma accelerator experiments at DESY. These observations are further analyzed using Monte Carlo and particle-in-cell simulations to improve the understanding of radiation generation mechanisms in plasma accelerators. The results identify key aspects that must be addressed when scaling plasma accelerator technology toward higher average powers and advancing it toward reliable operation in practical applications. %While the present study does not aim at a full quantitative prediction of absolute dose rates, it establishes the dominant radiation mechanisms and loss locations.

\section{Radiation Generation in Electron Accelerators}
\subsection{Radiation Generation Mechanisms}
The description of radiation generation mechanisms in electron accelerators presented in this section is based primarily on established treatments in IAEA TRS~188 \cite{Swanson1979RadiologicalAccelerators} and NCRP Report No.~144 \cite{NationalCouncilonRadiationProtectionandMeasurementsNCRP2003RadiationFacilities}, which provide comprehensive guidance for radiation protection at accelerator facilities.

In electron accelerators, ionizing radiation is produced predominantly through interactions of accelerated electrons with matter, such as beamline components, diagnostics, collimators, or beam dumps. The dominant mechanism is bremsstrahlung, generated when electrons are decelerated in the electromagnetic fields of atomic nuclei. The resulting photon spectrum is continuous and extends up to the full kinetic energy of the incident electrons, and typically constitutes the primary source of radiation exposure before shielding.

Bremsstrahlung photons generate secondary radiation through Compton scattering and pair production, producing electrons and positrons that further contribute to local dose deposition. At sufficiently high photon energies, photonuclear reactions occur, leading to neutron emission. Photoneutron production typically becomes relevant above material-dependent threshold energies of approximately \SI{8}{\mega\electronvolt} to \SI{10}{\mega\electronvolt}. Although neutron yields are lower than photon yields, neutrons are particularly important for radiation protection due to their high penetration depth and biological effectiveness, and therefore play a key role in shielding design and area classification.

Once a significant neutron component is present, activation of surrounding materials becomes an additional concern. Photon- and neutron-induced nuclear reactions in structural materials, beamline components, and shielding can produce radioactive isotopes, resulting in residual dose rates after beam operation has ceased. Activation is therefore particularly relevant for maintenance planning and access control.

At higher electron energies, additional radiation components become relevant. For energies of several hundred MeV and above, electromagnetic showers develop more fully, increasing both the spatial extent and complexity of radiation fields. Muons can be produced via direct lepton pair production by high-energy photons (above approximately \SI{200}{\mega\electronvolt}) and via photonuclear interactions followed by meson decay. Their yield increases with electron energy and beam power. Due to their long lifetime and high penetration capability, muons can contribute to radiation fields at large distances from primary beam-loss locations and may pose challenges for shielding design in high-energy or high-power facilities.

Overall, as emphasized in both reports \cite{Swanson1979RadiologicalAccelerators,NationalCouncilonRadiationProtectionandMeasurementsNCRP2003RadiationFacilities}, the relative contributions of different radiation components depend strongly on electron energy, beam power, loss geometry, and surrounding materials. Consequently, even comparatively low-energy accelerators can generate complex radiation fields that must be carefully assessed for both personnel safety and equipment protection.

\subsection{Scaling of Radiation Fields in Electron Accelerators}

\begin{figure}[htb]
\centering
\begin{subfigure}[t]{.48\textwidth}
  \centering
  \includegraphics[width=\textwidth]{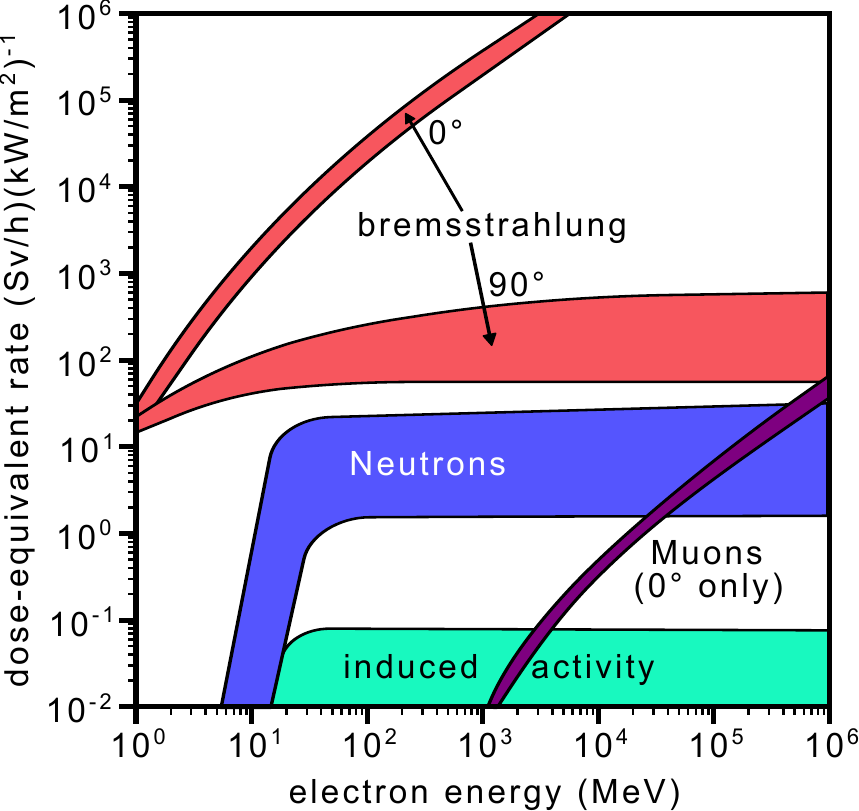}
  \caption{Schematic scalings of the dominant radiation components generated in electron accelerators as a function of electron energy \cite{Swanson1979RadiologicalAccelerators}.}
  \label{fig:trs188_fig6}
\end{subfigure}
\hspace{2mm}
\begin{subfigure}[t]{.48\textwidth}
  \centering
  \includegraphics[width=\textwidth]{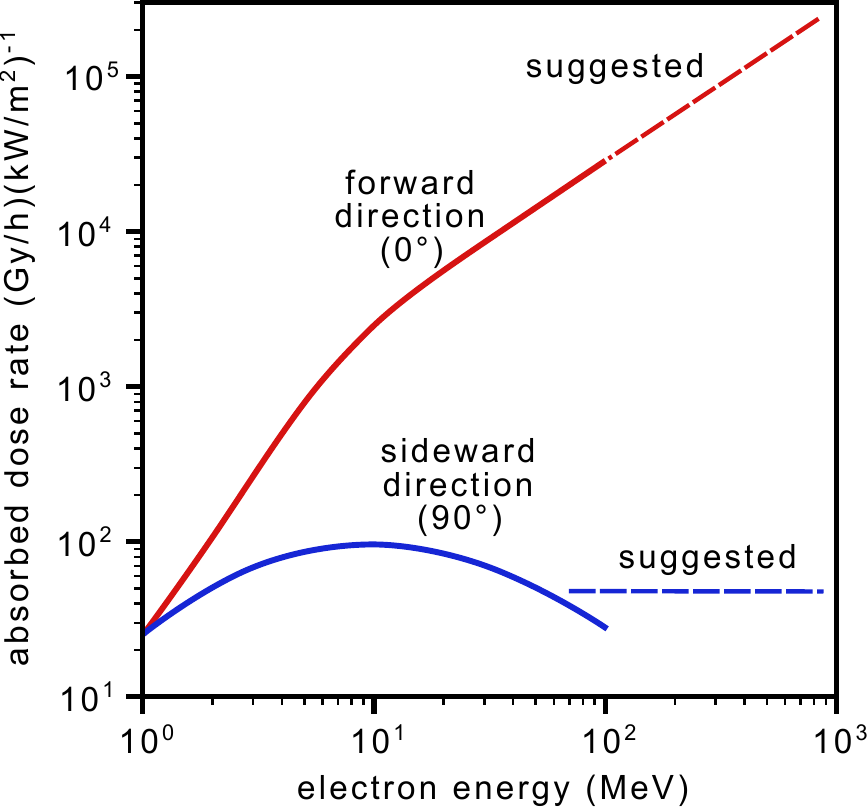}
  \caption{Bremsstrahlung dose-equivalent scaling as a function of electron energy for different emission geometries \cite{Swanson1979RadiologicalAccelerators}.}
  \label{fig:trs188_fig17}
\end{subfigure}%
\caption{Radiation generation mechanisms in electron accelerators as a function of electron energy. Both subfigures are redrawn as vector graphics based on Fig.~6 and Fig.~17 of IAEA TRS~188~\cite{Swanson1979RadiologicalAccelerators}.}
\label{fig:TRS_Scalings}
\end{figure}

The scaling of radiation fields in electron accelerators is summarized in IAEA TRS~188 \cite{Swanson1979RadiologicalAccelerators}. Of particular relevance are dose-equivalent scaling laws as a function of electron energy, beam power, and emission geometry. Figures~6 and~17 of the report are recreated in Fig.~\ref{fig:TRS_Scalings} and provide a reference framework for interpreting radiation generation in plasma accelerators.

Both figures present dose-equivalent or absorbed dose rates normalized to the incident beam power and evaluated at a reference distance of \SI{1}{\meter}. The quantities therefore correspond to dose rates per unit beam power at unit distance. This normalization allows scaling to arbitrary operating conditions by multiplying with the beam power and applying an inverse-square dependence on distance. While this point-source approximation strictly applies only to localized, unshielded beam losses, it provides a useful framework for many radiation protection applications.

The radiation components shown in Fig.~\ref{fig:trs188_fig6} are assumed to originate from a localized beam-loss point (e.g.\ a beam dump or structural component). The figure presents the relative contributions of different radiation components to the dose equivalent as a function of electron energy, based on a qualitative compilation of dose-equivalent rates per unit beam power. Bremsstrahlung photons dominate over the full energy range, while neutron production becomes relevant at several tens of MeV and muon contributions appear at GeV-scale energies. The width of the bands reflects variability due to material, geometry, and interaction conditions and is based on a point-source approximation with inverse-square scaling.

An important observation is that, for many radiation components such as sideways-emitted bremsstrahlung and photoneutrons, the dose equivalent shows only a weak dependence on electron energy once production thresholds are exceeded and electromagnetic showers are fully developed. In these regimes, the dose scales predominantly with beam power rather than beam energy, which is particularly relevant for high-repetition-rate linear and plasma accelerators.

Figure~\ref{fig:trs188_fig17} presents the absorbed dose-equivalent from thick-target bremsstrahlung as a function of electron energy for different emission geometries. The ``suggested'' curves represent envelope estimates of experimental data accounting for variations in target composition and measurement conditions. The sideways ($90^\circ$) component, relevant for shielding of experimental halls, shows significant variability and depends strongly on geometry. The forward ($0^\circ$) direction exhibits a stronger dependence on electron energy.

This difference arises from electromagnetic shower development. At high energies, radiation is strongly forward-peaked due to Lorentz boosting, while successive interactions broaden the angular distribution. As a result, a persistent sideways component remains, particularly at lower energies and later stages of the shower. Consequently, even highly forward-directed beams produce significant lateral dose contributions.

Electrons emitted at intermediate angles contribute dose rates between the forward and sideways limits, forming a transition region between the two geometries.

Overall, the scaling laws summarized in TRS~188 show that significant radiation doses can occur already at relatively low electron energies and that beam power, rather than beam energy, is often the dominant parameter.

\subsection{Implications for Plasma Accelerators}
The general scaling laws discussed above have several important implications for plasma accelerators. In contrast to conventional radio-frequency linacs, plasma accelerators typically produce electron beams with relatively large energy spreads and significant divergence. As a result, the angular distribution of electrons plays a crucial role in radiation generation, and dose production cannot be assessed solely based on on-axis beam parameters.

In particular, low-energy electrons can contribute disproportionately to radiation dose. Electrons in the low-energy tail of the spectrum, for example originating from continuous ionization injection or from down-ramp injection at the exit of the plasma, produce bremsstrahlung efficiently in the energy range around \SI{10}{\mega\electronvolt}, where sideways dose-equivalent production is maximal. These electrons therefore cannot be neglected in radiation safety assessments, even if they do not contribute to the usable high-energy beam.

Furthermore, electrons emitted from the plasma at large angles with respect to the beam axis lead to distributed beam losses and enhanced lateral radiation. While such electrons have been reported previously for laser--plasma accelerators \cite{Kaganovich2008ObservationWakefield,Yang2017ThreeQuestion,Garrett2025ExcitationAcceleration}, they are typically not accounted for in radiation protection assessments. These loss patterns can produce significant dose rates in experimental halls and near sensitive equipment, particularly in compact layouts with limited shielding distances, and therefore need to be considered at high average powers.

In conventional accelerator design, beam losses are typically localized to a small number of well-defined components (``points''), such as beam dumps or collimators. As emphasized in IAEA TRS~188, ``It is good practice to have as few such points as possible in order to simplify the shielding requirements'' \cite{Swanson1979RadiologicalAccelerators}. This allows radiation sources to be treated as discrete locations and facilitates optimized shielding strategies.

In plasma accelerators, however, this assumption is often difficult to fulfil. Due to intrinsic beam divergence, energy spread, and injection mechanisms, beam losses can occur over extended regions and at multiple locations along the beamline. This distributed loss pattern complicates shielding design and leads to more complex radiation fields. It also increases the likelihood of radiation exposure of nearby components, including electronics and magnetic systems, potentially resulting in performance degradation or failure. These aspects will be discussed in the following section, while an example of the increased complexity is described in Sec.~\ref{sec:SimandMeasatKALDERA}.

\section{Radiation Damage to Electronics and Magnets}
\label{sec:raddamage}
In addition to personnel safety, ionizing radiation can affect accelerator components, particularly electronic systems and magnetic elements. These effects must be considered when designing shielding concepts, defining component placement, and assessing operational limits in plasma accelerator facilities.

Electronic components are susceptible to both cumulative and stochastic radiation effects \cite{George2019AnElectronics}. Total ionizing dose (TID) leads to long-term degradation of semiconductor devices, with functional failure of commercial electronics typically occurring in the range of \SI{0.1}{\kilo\gray} to \SI{1}{\kilo\gray}, while radiation-hardened components can tolerate significantly higher doses. In addition, single-event effects induced by individual high-energy particles can cause transient or permanent malfunctions even at low integrated dose, which is particularly relevant in the mixed radiation fields of electron accelerators.

Magnetic components, especially permanent magnets, can also degrade under irradiation. Typical flux losses of about 1\% are observed at doses of \SI{10}{\kilo\gray}--\SI{100}{\kilo\gray} \cite{Skupin2008UndulatorFLASH,Bizen2016Radiation-inducedMagnets}, although localized and non-linear damage mechanisms can lead to significantly larger degradation under certain conditions \cite{Bizen2016Radiation-inducedMagnets}.

These effects are particularly relevant for plasma accelerators due to their compact layouts and potentially distributed beam losses. Experience at DESY shows that radiation-induced degradation of components is a general feature of accelerator operation, affecting both diagnostics close to beamlines and technical infrastructure inside the accelerator tunnels. 
\begin{comment}
    For the plasma accelerator FLASHForward \cite{DArcy2019FLASHForward:Applications}, which operates in a shared tunnel with FLASH2, radiation exposure of nearby undulator magnets is a key consideration. To mitigate this risk, the average beam power is conservatively limited to approximately \SI{20}{\watt} to prevent any damage to the permanent magnets of the FLASH2 undulator section. Despite this, FLASHForward represents the highest average beam power plasma accelerator experiment at DESY and requires dedicated shielding of sensitive components. At the same time, it operates reliably within controlled radiation conditions, and radiation damage to electronics has not emerged as a dominant operational limitation.
\end{comment}

For the plasma accelerator FLASHForward \cite{DArcy2019FLASHForward:Applications}, which operates in a shared tunnel with FLASH2, radiation exposure of nearby undulator magnets and other accelerator components is a key consideration. The FLASHForward beamline is located in close proximity to the undulator section of FLASH2, such that secondary radiation from the plasma stage or from disposal of the spent driver bunch can reach the permanent magnets, which are susceptible to radiation-induced demagnetization. To mitigate this risk, the average beam power is conservatively limited to \SI{20}{\watt} (compared to tens of kW for FLASH2). This limit is not derived from a dedicated radiation damage calculation, but represents a practical operational constraint that balances the desire to perform plasma acceleration experiments with high statistics, while ensuring the reliable operation of, in this case, a photon generation beamline and its ancillary components.

Despite this limitation, FLASHForward currently constitutes the highest average beam power plasma accelerator experiment currently operated at DESY and therefore requires dedicated shielding of sensitive components such as cameras near spectrometers and vacuum pumps in the vicinity of the beamline. This proactive design enables reliable operation and limits radiation damage to sensitive equipment.

Radiation effects on electronic and magnetic systems thus represent a fundamental design constraint for future plasma accelerator user facilities such as EuPRAXIA \cite{Assmann2020EuPRAXIAReport}, and become even more critical for high-power plasma-based concepts such as HALHF \cite{Foster2023AAcceleration}, where beam powers in the megawatt range are envisaged. In such facilities, radiation protection, shielding design, component placement, and robust beam-dump concepts must be integrated into the accelerator design from the earliest conceptual stage to ensure reliable long-term operation.

\section{Simulation and Measurement Studies at KALDERA}
\label{sec:SimandMeasatKALDERA}
\subsection{The KALDERA Laser Plasma Accelerator at DESY}
\begin{figure}[htb]
  \centering
  \includegraphics[width=0.9\linewidth]{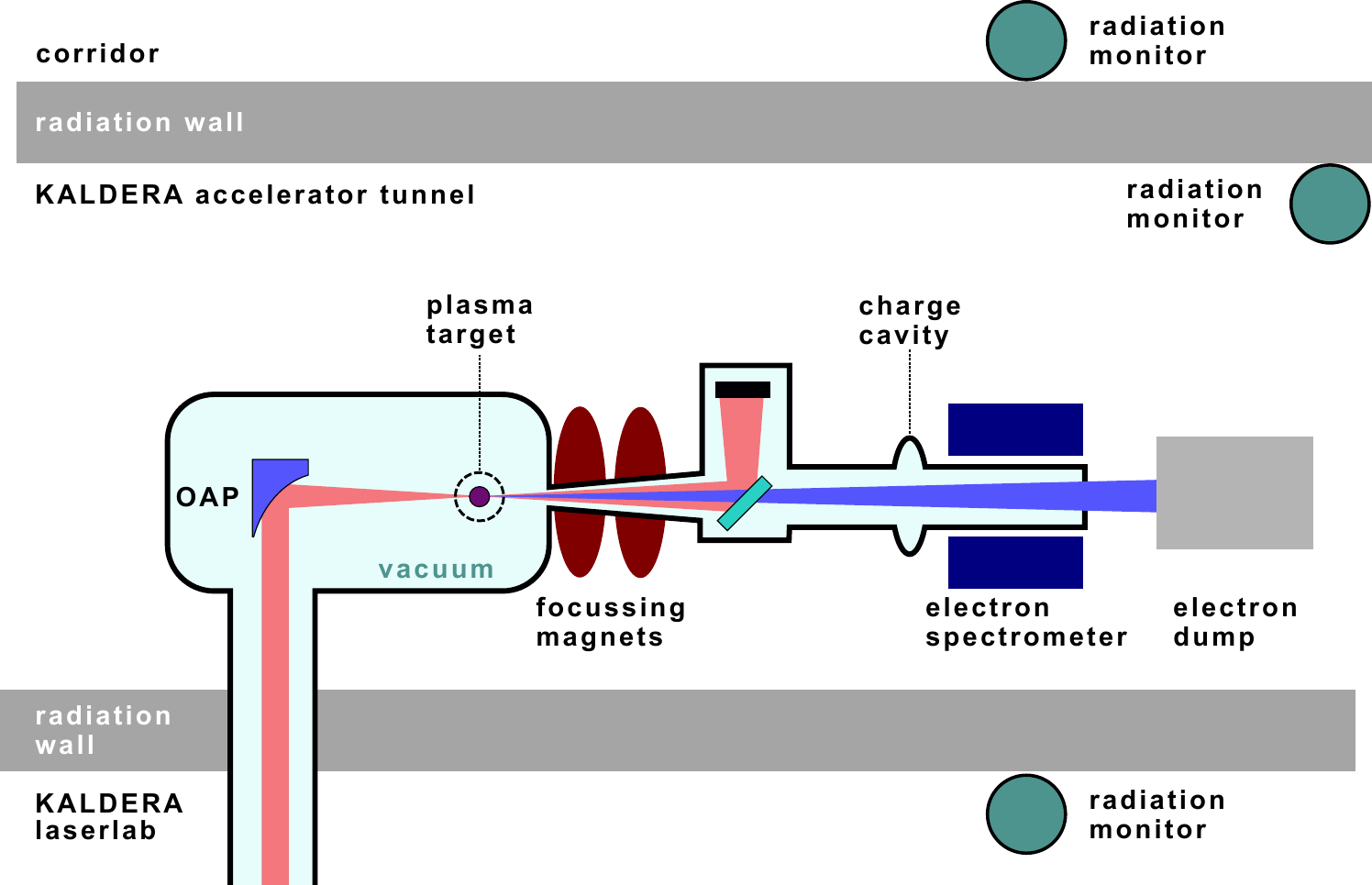}
  \caption{Simplified schematic overview of the KALDERA setup and the position of radiation monitors in- and outside the accelerator tunnel (not to scale).}
  \label{fig:ExpSetupKaldera}
\end{figure}

The KALDERA laser plasma accelerator is an experimental facility at DESY designed to investigate laser-driven plasma acceleration at high average power \cite{ManuelKirchen2025FirstKALDERA}. The setup utilises a high-power, ultrashort-pulse laser system currently operating at \SI{100}{Hz}, with an upgrade to \SI{1000}{Hz} foreseen, to generate relativistic electron bunches over acceleration lengths of only a few centimeters. In the first development phase, the KALDERA laser delivers up to \SI{500}{mJ} of pulse energy to drive the MAGMA laser-plasma accelerator. A schematic overview of the experimental setup is shown in Fig.~\ref{fig:ExpSetupKaldera}. Although KALDERA is a laser-driven plasma accelerator, its radiation characteristics are representative of high-average-power plasma accelerators more generally.

During its initial commissioning phase, KALDERA routinely produces electron beams with energies in the range of \SIrange{100}{150}{MeV} and bunch charges around \SI{20}{pC} at \SI{100}{Hz} repetition rate \cite{ManuelKirchen2025FirstKALDERA}. The generated beams exhibit characteristic features of laser plasma acceleration, including percent-level energy spread, mrad-level beam divergence, and significant shot-to-shot variations in beam parameters. Electron beam diagnostics based on a cavity-based charge monitor \cite{Bohlen2025NoninvasiveAccelerators} and a magnetic spectrometer provide reliable measurements of the energy spectrum and total charge for particles transported along the beamline. However, these diagnostics are inherently limited to electrons that are captured and guided into the beamline and thus reach the charge cavity and the spectrometer. Electrons exiting the plasma target at large angles are not intercepted by these detectors and cannot be measured directly.

From a radiation safety perspective, these beam characteristics make KALDERA a representative test case for studying radiation generation in plasma accelerators at increasing average powers. The compact layout of the experimental area, together with potential distributed beam losses, results in complex radiation fields that differ from those in conventional radio-frequency linacs. Therefore, radiation monitoring at KALDERA extends beyond personnel protection outside the accelerator, with a dedicated monitor installed inside the tunnel to enable detailed characterization of radiation fields during operation. As such, KALDERA provides a valuable platform for benchmarking Monte Carlo and particle-in-cell simulations against measurements and for developing radiation protection concepts tailored to plasma-based accelerator facilities.
%From a radiation safety perspective, these beam characteristics make KALDERA a representative test case for studying radiation generation in plasma accelerators at increasing average powers. The compact layout of the experimental area, together with the potential for distributed beam losses, results in complex radiation fields that differ from those encountered in conventional radio-frequency linacs. Therefore, radiation monitoring at KALDERA extends beyond personnel protection outside the accelerator, with a dedicated radiation monitor installed inside the tunnel to allow detailed characterization of the radiation fields produced during operation. As such, KALDERA provides a valuable platform for benchmarking Monte Carlo and particle-in-cell simulations against measurements and for developing radiation protection concepts tailored to plasma-based accelerator facilities.

\subsection{Initial Monte Carlo Simulations}
Prior to first experimental operation of KALDERA, Monte Carlo simulations were performed using FLUKA \cite{Ferrari2005FLUKA2005,Bohlen2014TheApplications} to obtain an initial estimate of the expected radiation fields. The simulations incorporated the geometry of the accelerator tunnel and a simplified beam-dumping concept, including the bending magnet and the electron dump, assuming efficient transport of the accelerated electrons into the dump.

The simulations were carried out for electron beams with an energy of \SI{100}{MeV}, a 5\% rms energy spread and a bunch charge of \SI{100}{pC} at a repetition rate of \SI{100}{Hz}. These parameters correspond to an optimistic estimate of the first operational beam conditions in terms of average beam power. The resulting radiation fields therefore represent an upper-bound estimate for the initial commissioning phase.

\begin{figure}[H]
  \centering
  \includegraphics[width=0.75\linewidth]{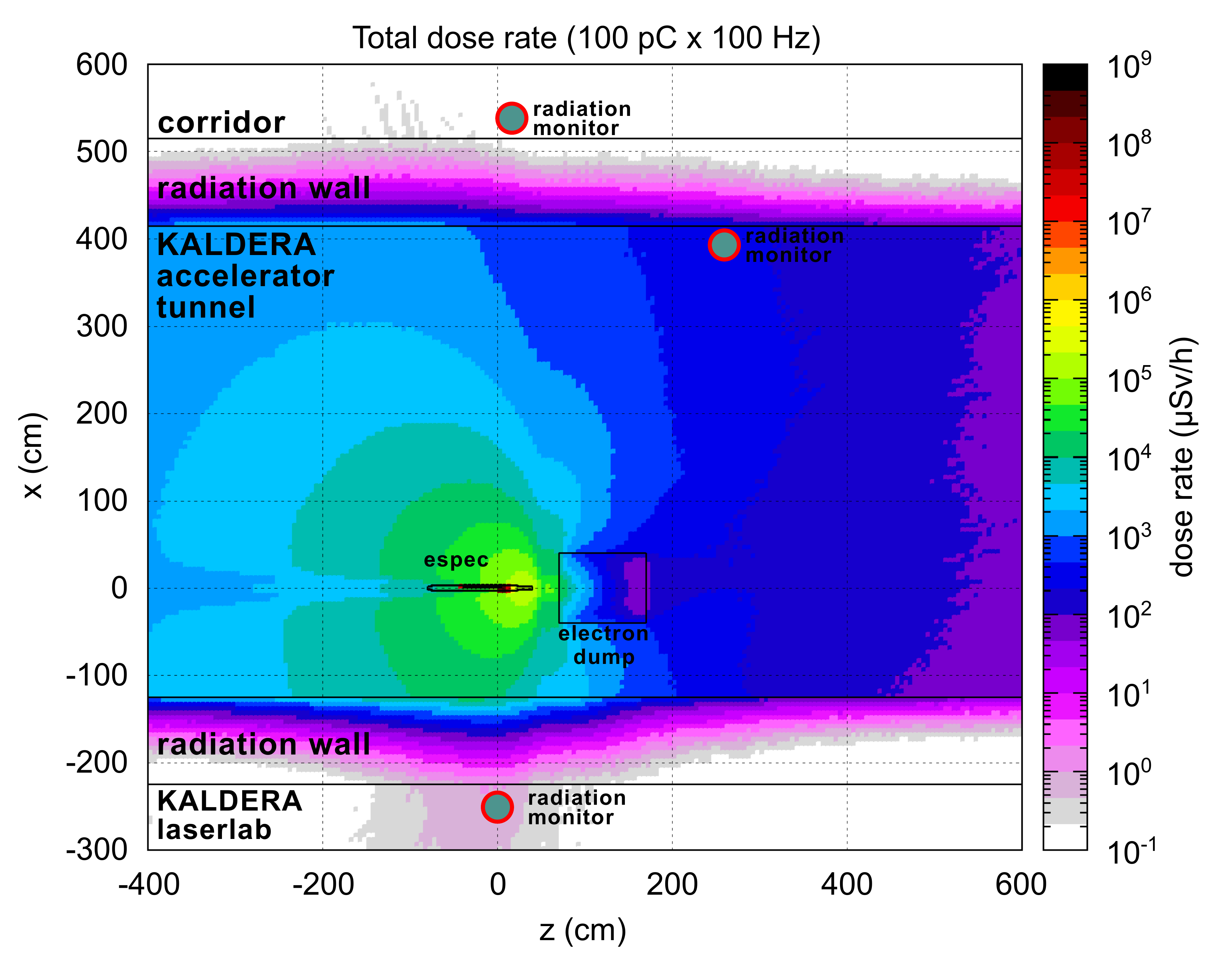}
  \caption{Top-view FLUKA simulation of radiation dose fields (electrons, photons, and neutrons) in the horizontal plane at beamline level for simplified KALDERA beam parameters. The tunnel geometry and a beam-dump setup are included and radiation monitor positions are indicated.}
  %Top-view FLUKA simulation of radiation dose fields of all particles (electrons, photons and neutrons) in the horizontal plane at beamline level for simplified KALDERA beam parameters. The tunnel geometry and simplified beam-dumping setup are included. The position of radiation monitors is indicated for reference.}
  \label{fig:FlukaInitial}
\end{figure}

The simulated dose distributions are shown in Fig.~\ref{fig:FlukaInitial}. For orientation, the position of the radiation monitors is indicated in the figure. Under the assumed conditions, the simulations predict radiation levels consistent with localized beam losses at the dump and provide a baseline expectation for comparison with subsequent radiation measurements.

\subsection{Radiation Measurements}
\begin{figure}[H]
\centering
\begin{subfigure}[t]{.48\textwidth}
  \centering
  \includegraphics[height=7cm]{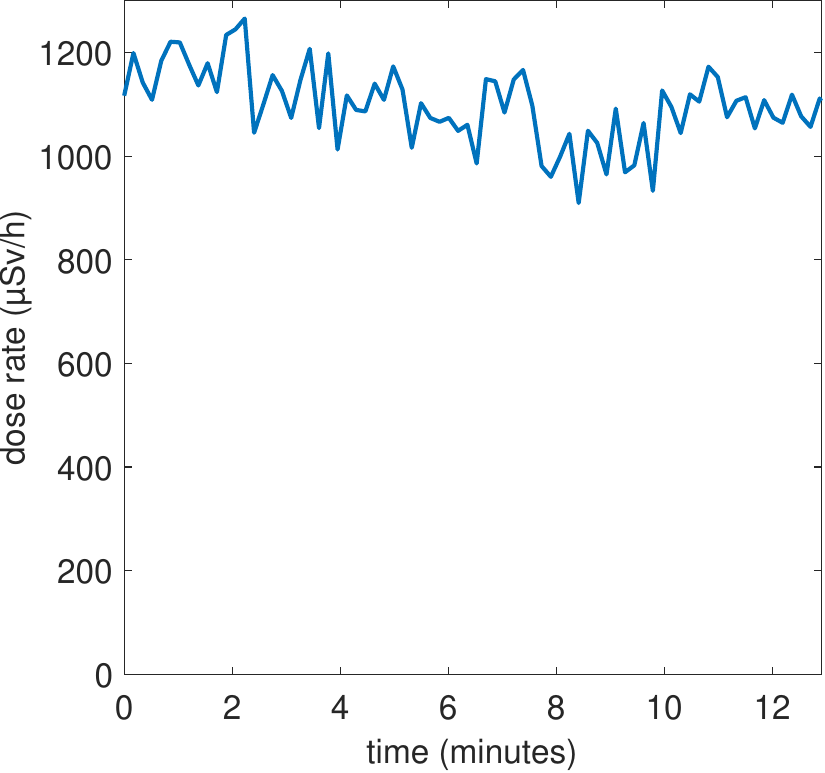}
  \caption{Measured dose rate (saturation corrected) for the electromagnetic dose component during electron acceleration as reported in \cite{ManuelKirchen2025FirstKALDERA}.}
  \label{fig:dose_stability}
\end{subfigure}%
\hspace{4mm}
\begin{subfigure}[t]{.48\textwidth}
  \centering
  \includegraphics[height=7cm]{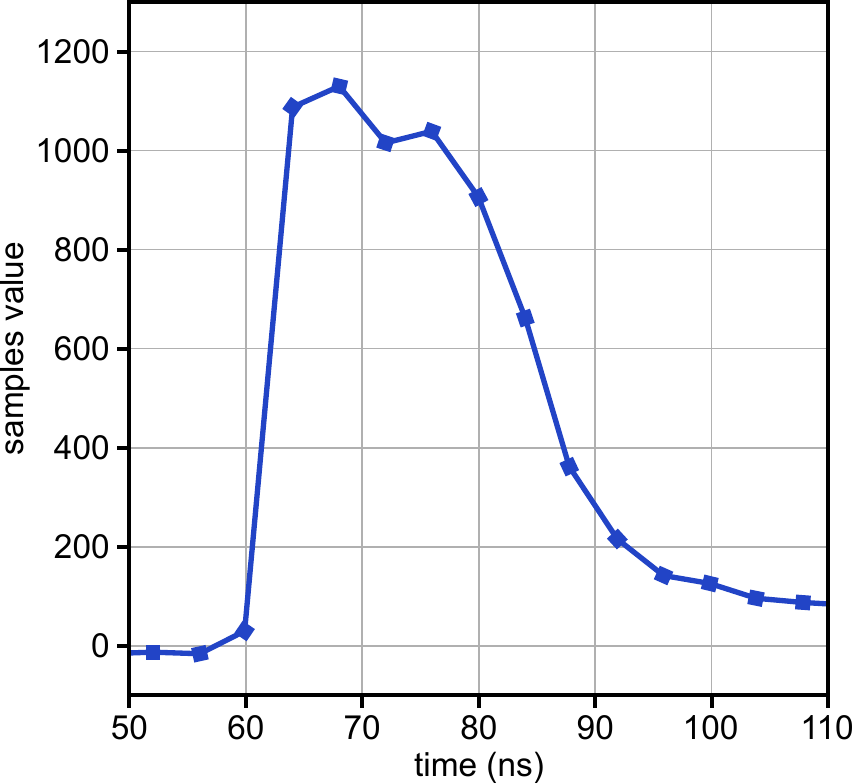}
  \caption{Time-resolved scintillator signal, showing a double-peak structure attributed to radiation originating from different locations along the beamline.}
  \label{fig:RadMeasurementTiming}
\end{subfigure}
\caption{Measurements with the LB6419 detector inside the KALDERA accelerator tunnel during electron acceleration}
\label{fig:dosemeasurements}
\end{figure}

Radiation measurements at KALDERA were performed using Berthold LB6419 radiation monitors, a detector system developed at DESY for measurements of pulsed radiation from particle accelerators \cite{Klett2010AFields}.  Results are shown in Fig.~\ref{fig:dosemeasurements}. The LB6419 is capable of independently measuring photon and neutron dose rates by combining a plastic scintillator for photons with a moderated $^3$He proportional counter for neutrons, allowing a separation of the dominant radiation components in mixed radiation fields.

During the initial operation of KALDERA, radiation levels measured at the detector in the accelerator tunnel exceeded those predicted by Monte Carlo simulations, despite the measured electron beam power being lower than assumed in the simulations. The discrepancy was most pronounced for the electromagnetic dose component. After applying a saturation correction to account for the detector’s non-linear response at high dose rates, dose levels of approximately \SIrange{1000}{1500}{\micro\sievert\per\hour} were inferred during beam optimisation. For the acceleration reported in \cite{ManuelKirchen2025FirstKALDERA}, a value of \SI{1097 \pm 73}{\micro\sievert\per\hour} was obtained (see Fig.~\ref{fig:dose_stability}), compared to simulated values of only \SI{60}{\micro\sievert\per\hour}. The correction was derived from dedicated measurements at other DESY accelerators with well-defined beam parameters and benchmarked at dose rates up to two orders of magnitude higher, and is therefore considered to be robust.

In addition, the scintillator signal exhibited a characteristic double-peak structure, indicating two distinct radiation contributions separated by a few nanoseconds, as shown in Fig.~\ref{fig:RadMeasurementTiming}. This timing difference corresponds to an effective path-length difference of approximately \SIrange{2}{3}{m}.

\begin{comment}
    \begin{figure}[H]
  \centering
  \includegraphics[width=0.48\textwidth]{Figures/ADC_2peak.pdf}
  \caption{Time-resolved scintillator signal measured with the LB6419 detector during KALDERA operation, showing a double-peak structure attributed to radiation originating from different locations along the beamline.}
  \label{fig:RadMeasurementTiming}
\end{figure}
\end{comment}

This observation suggests an additional radiation source close to the plasma target, superimposed on the radiation originating from the electron dump. The path-length difference between radiation traveling directly from the plasma source to the detector and radiation produced at the dump (after transport from the source) is consistent with the measured timing separation. To investigate this hypothesis and to better understand the underlying beam-loss mechanisms, further studies were carried out using detailed particle-in-cell simulations of the plasma acceleration stage.

\subsection{Particle-In-Cell Simulations}
Particle-in-cell simulations were performed using FBPIC in a boosted frame~\cite{Lehe2016AAlgorithm} to reproduce the electron generation observed during the initial commissioning phase of the experiment. The simulation setup was chosen to match the experimental conditions and was found to reproduce the formation of the primary electron bunch with an energy of \SI{105}{MeV} propagating along the direction of the driving laser pulse. The detailed physical conditions can be found in Ref.\,\cite{ManuelKirchen2025FirstKALDERA}. The simulation ran with 40 cells per laser wavelength longitudinally, and 4 cells per wavelength radially. Hydrogen was fully pre-ionized, and Argon was ionized up to the 8th level. 

\begin{figure}[H]
\centering
  \centering
  \includegraphics[width=\textwidth]{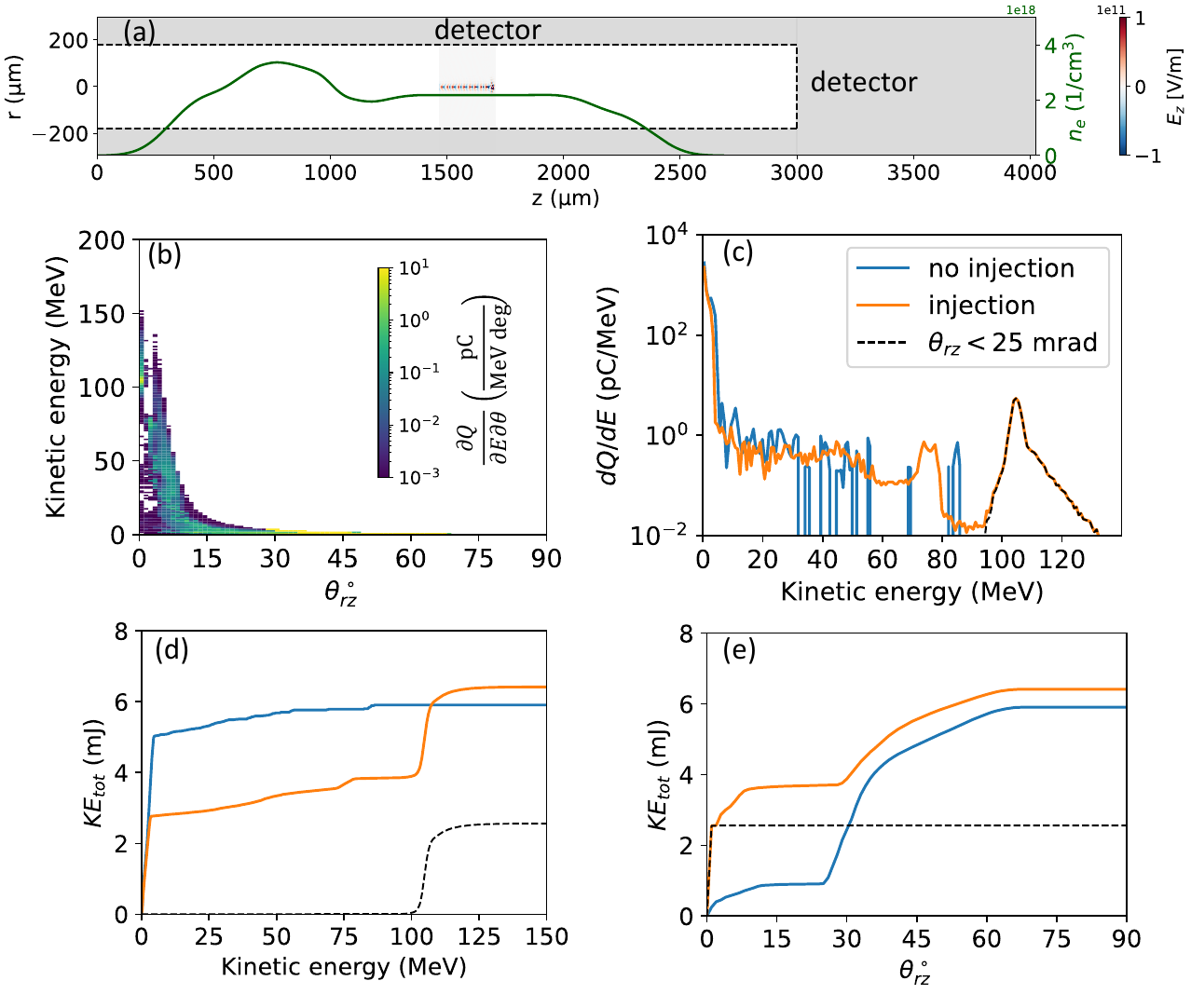}
  \caption{FBPIC simulation of electrons escaping the plasma. (a) Simulation setup and virtual detectors; (b) Energy-angle distribution of electrons in the case with injection; (c) Spectra of emitted electrons; (d) cumulative kinetic energy spectrum. The value at e.g. 100~MeV represents  the total kinetic energy of all electrons with an energy lower or equal to 100~MeV. (e) Same as (d), with the abscissa representing the electron propagation angle with respect to propagation direction.}
  \label{fig:pic1}
\end{figure}

In addition to standard diagnostics for the forward-propagating beam, virtual detectors were implemented to record all electrons exiting the plasma, both in the forward direction (in the simulation box after the final downramp) and at large angles relative to the laser axis (with radius $r>180~\mu m$). The simulated laser pulse has an energy of \SI{484}{\milli\joule} and a duration of \SI{30}{\femto\second} (FWHM). The plasma profile has a length of \SI{2.7}{\milli\meter} and a transverse width of \SI{150}{\micro\meter}. The longitudinal plasma density profile follows the shape shown in Fig.~\ref{fig:pic1}a, while the transverse profile is assumed to be uniform. These parameters were chosen to closely reproduce the experimental conditions at KALDERA. To capture electrons traveling at large angles, a simulation box longer than typically used was employed, as illustrated in Fig.~\ref{fig:pic1}a. This extended diagnostic configuration enabled a comprehensive characterization of electron emission over a wide angular range, allowing the detection of particles emitted at large angles, as previously observed in laser–plasma accelerators \cite{Kaganovich2008ObservationWakefield,Yang2017ThreeQuestion,Garrett2025ExcitationAcceleration}, and as is also currently being investigated for beam-driven plasma accelerators \cite{I.Najmudin2025TemperatureAccelerator,J.Cowley2025DiagnosticsFlow}.

The simulations reveal that, in addition to the main electron bunch traveling along the laser propagation direction, a substantial amount of charge is generated at large emission angles of up to approximately \SI{70}{\degree}. The correlation between emission angle and kinetic energy is shown in Fig.~\ref{fig:pic1}b. The spectrum shown on Fig.~\ref{fig:pic1}c clearly shows the large amount of few-\si{\mega\electronvolt} electrons at large angles, as well as the forward-directed monoenergetic bunch. The total charge reaches values of up to \SI{3}{\nano\coulomb}, while the charge above 1~MeV (5~MeV) is \SI{1}{\nano\coulomb} (\SI{55}{\pico\coulomb}).

An analysis of the energy distribution shows that, while the forward-propagating bunch dominates transport along the beamline, the total energy carried by electrons emitted at large angles is comparable to that of the main bunch. Approximately 40\% of the total electron energy generated in the interaction is contained in the forward-directed beam, with the remaining fraction distributed among electrons emitted at large angles, as shown in Fig.~\ref{fig:pic1}d-e. In the simulations, low-energy electrons (on the order of a few MeV) are predominantly emitted at large angles and exhibit a broad angular distribution, with a pronounced spread in the $y$ direction, i.e., orthogonal to the laser polarization axis ($x$). This behavior is consistent with previous studies on large-angle electron emission in laser–plasma accelerators \cite{Yang2017ThreeQuestion}. Furthermore, the simulations confirm that variations in the laser peak intensity significantly influence the transverse distribution of the emitted electrons in the $x$–$y$ plane.

Additionally, a working point with \emph{no injection} was investigated. While the first working point is representative of a successful production shot, the one without injection is typical of the tuning phase of the accelerator. In practice, simulating this was achieved by artificially turning off ionization of the argon dopant above the 8th level in the simulation, which inhibits trapping while keeping all conditions the same. In particular, the laser pulse evolution and initial electron density was the same in both cases.

The properties of ejected electrons for this case are shown on Fig.~\ref{fig:pic1}c-e. The total energy of electrons is comparable in both cases (\SI{6}{mJ} vs. \SI{6.3}{mJ}), but the distributions differ: in the absence of injection, a vast majority of electrons detected are below 5~MeV and above \SI{25}{\degree}. The electrons detected sideways are mostly sheath electrons that obtain a large transverse momentum by interacting with the back of the bubble. In the presence of a beam, its space-charge force slows down the sheath electrons via \emph{beam loading}, which could explain the lower ejected charge observed.

These results demonstrate substantial electron generation outside the nominal beam direction, consistent with previous observations in laser--plasma acceleration experiments. The simulations therefore support the hypothesis of an additional source of radiation close to the plasma target, contributing significantly to local radiation fields and potentially dominating radiation losses in the immediate vicinity of the source. A few simplifying assumptions were made in the simulation: electron dynamics in the walls of the plasma cell was not modeled, low-energy (non-relativistic) electrons were not all captured, the target is assumed to be pre-ionized, etc. Further studies would be needed for a better understanding of these electrons.

\subsection{Updated Monte Carlo Simulations}
To systematically investigate the impact of different electron source terms on the resulting radiation fields, three additional FLUKA simulations were performed using the electron distributions shown in Fig.~\ref{fig:pic1}c–e. Together with the initial simulation, this results in four cases that can be directly compared (see Tab.~\ref{tab:tabledoses}).

For the additional FLUKA simulations, the angular and energy distributions of electrons exiting the plasma, as obtained from the FBPIC simulations, were transferred to FLUKA without further parameterization. This ensures a self-consistent representation of both the high-energy core and the large-angle, low-energy electron components.

A simplified geometrical model of the plasma chamber and beamline was implemented to account for interactions of electrons with the immediate surroundings. The model includes the chamber volume and relevant structural boundaries but does not incorporate internal beam optics or detailed mechanical components. This level of detail was chosen to isolate the dominant radiation generation mechanisms associated with electron losses close to the plasma target while maintaining computational efficiency.

\begin{figure}[H]
\centering
\begin{subfigure}[t]{.49\textwidth}
  \centering
  \includegraphics[width=\textwidth]{Figures/FLUKA_case1_initial.pdf}
  \caption{Case 1: Initial idealized source term. Same as Fig.~\ref{fig:FlukaInitial}.}
  \label{fig:MC_Sim_OG_recreated}
\end{subfigure}
\hspace{0.2mm}
\begin{subfigure}[t]{.49\textwidth}
  \centering
  \includegraphics[width=\textwidth]{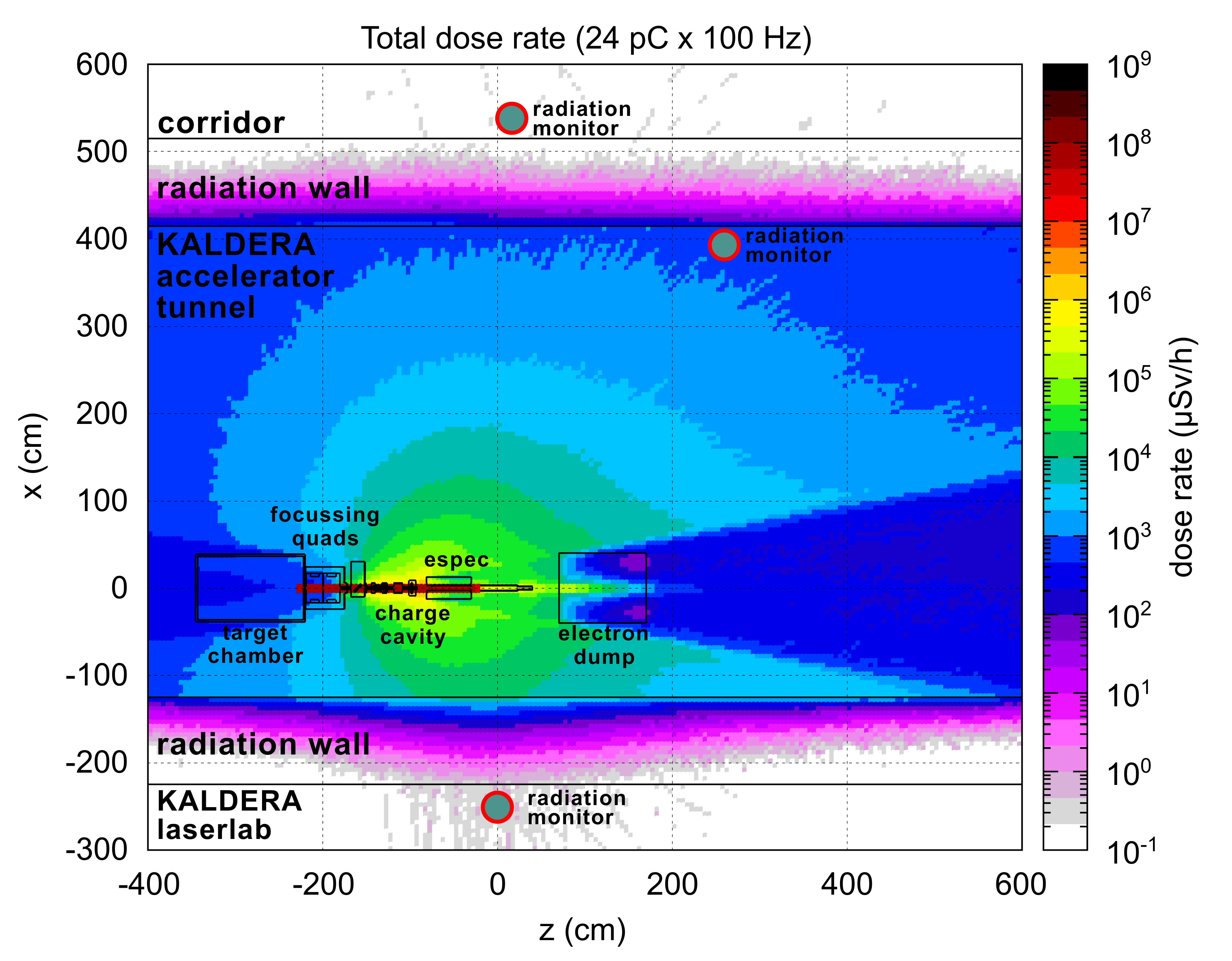}
  \caption{Case 2: Central ($\pm$\SI{25}{mrad}) PIC distribution (with injection).}
  \label{fig:MC_Sim_centreBeam}
\end{subfigure}%
\newline
\begin{subfigure}[t]{.49\textwidth}
  \centering
  \includegraphics[width=\textwidth]{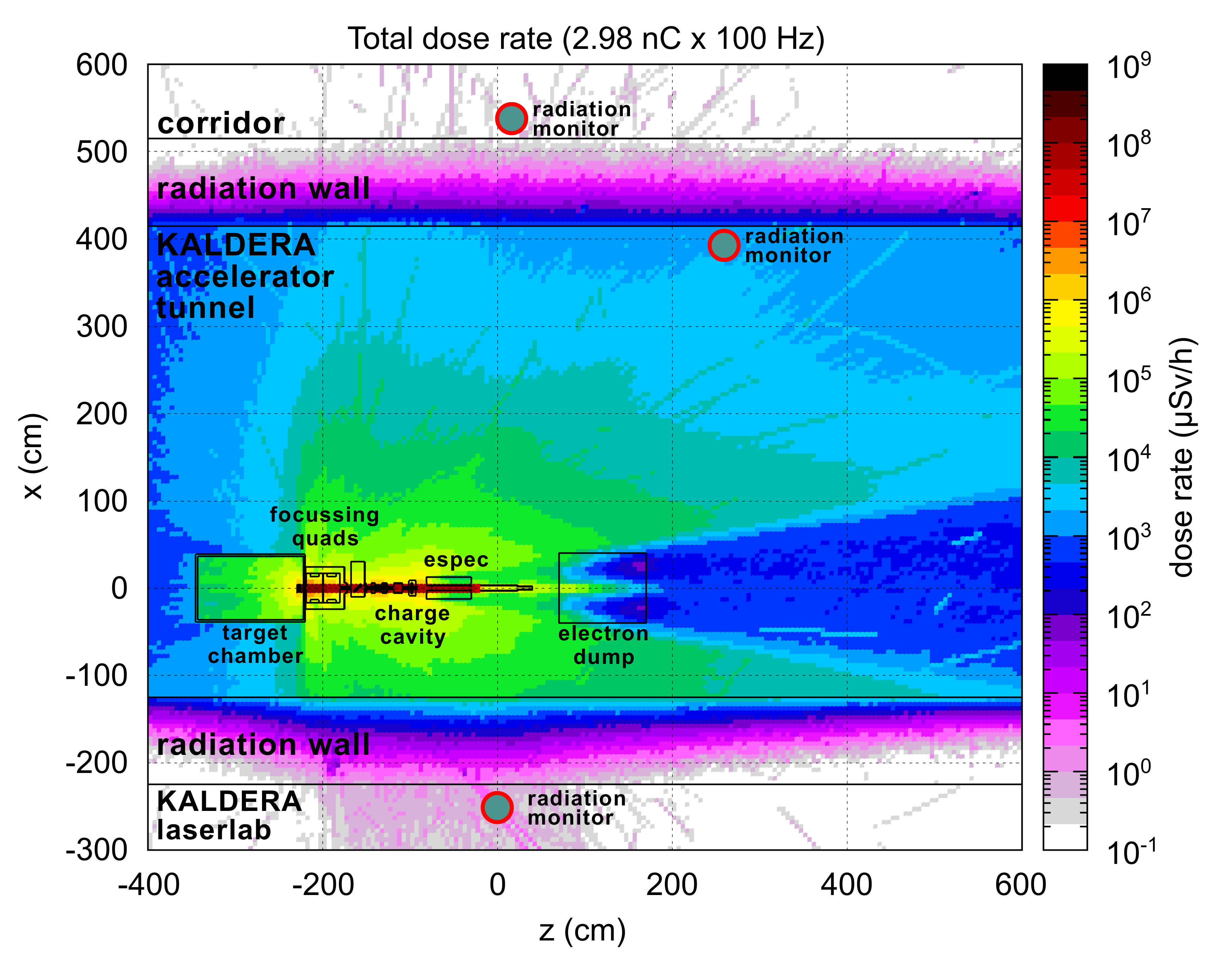}
  \caption{Case 3: Full PIC distribution (with injection).}
  \label{fig:MC_Sim_totalBeam}
\end{subfigure}
\hspace{0.2mm}
\begin{subfigure}[t]{.49\textwidth}
  \centering
  \includegraphics[width=\textwidth]{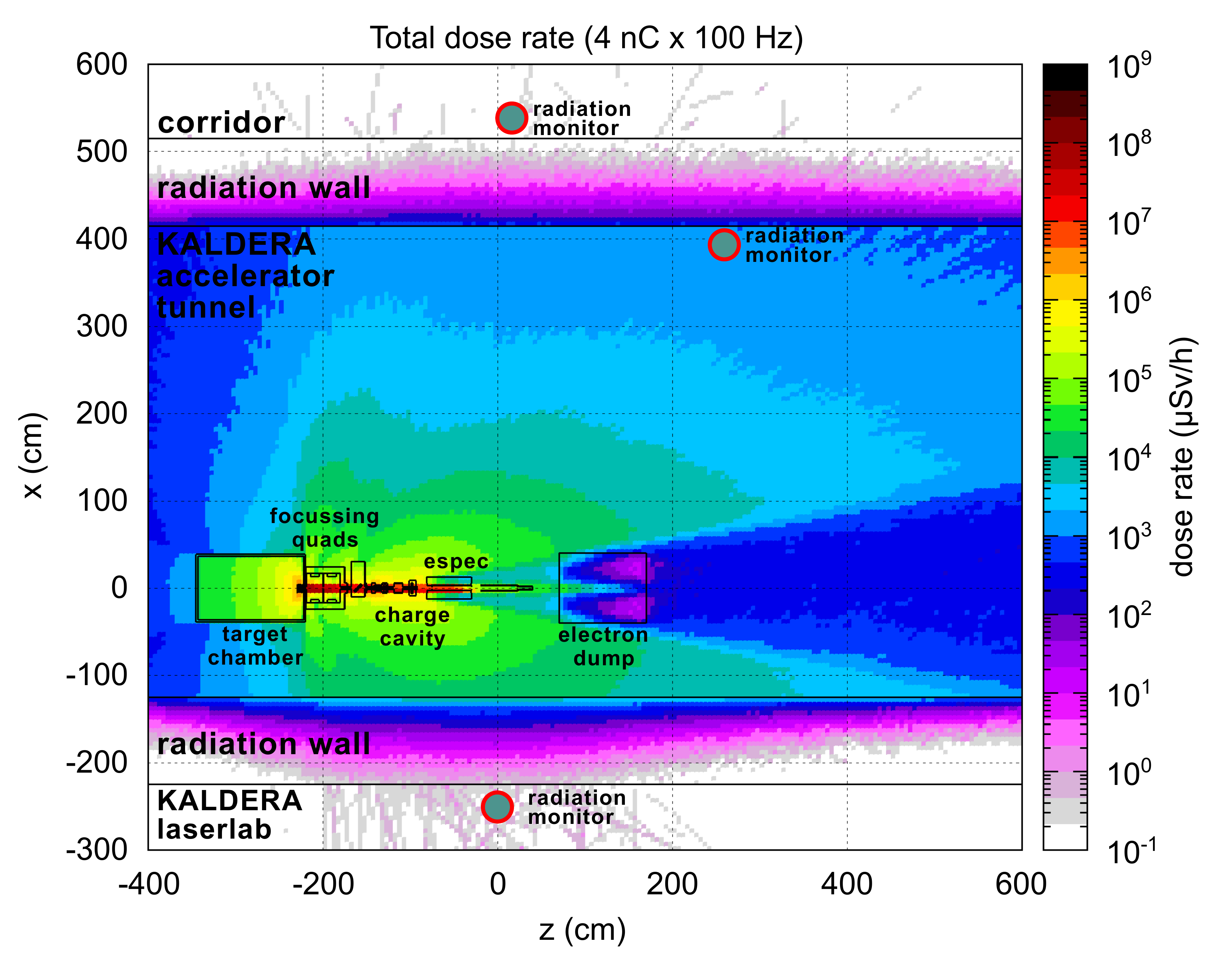}
  \caption{Case 4: Full PIC distribution (without injection).}
  \label{fig:MC_Sim_noAr}
\end{subfigure}
\caption{Comparison of radiation dose fields of all particles (electrons, photons and neutrons) for different source terms. All panels show top-view FLUKA simulations of radiation dose fields in the horizontal plane at beamline level.}
\label{fig:MC_comparison}
\end{figure}

Downstream beamline elements and the electron dump were retained from the initial model, allowing direct comparison between radiation generated near the plasma source and radiation originating from losses at the dump. The simulation results are summarized in Fig.~\ref{fig:MC_comparison}.

The initial configuration (Fig.~\ref{fig:MC_Sim_OG_recreated}) assumes a perfect beam transport to the dedicated dump. In contrast, restricting the source term to the central cone of the PIC distribution (\SI{\pm 25}{mrad}, Fig.~\ref{fig:MC_Sim_centreBeam}) already leads to substantially modified radiation fields. Although only about one quarter of the beam power (\SI{0.26}{W}) is assumed in these simulations, dose levels inside the accelerator tunnel increase compared to the idealized scenario. This enhancement arises from interactions of a small fraction of the (unfocussed) electron beam within the central cone with beamline components prior to the dump.

Including the full PIC electron distribution with injection (Fig.~\ref{fig:MC_Sim_totalBeam}) further increases the radiation fields. The total charge rises to approximately \SI{3}{nC}, corresponding to a beam power of about \SI{0.63}{W}. Despite remaining below the \SI{1}{W} assumed in the initial idealized simulation, the predominance of low-energy electrons emitted at large angles results in significantly elevated dose rates inside the tunnel, reaching several tens of millisieverts per hour near the beamline, exceeding those obtained in the idealized configuration. 
Dose levels behind the shielding (inside the laser laboratory) are comparable to those predicted in the initial simulation and higher than in the central-beam configuration. This demonstrates that electrons outside the nominal beam cone constitute a relevant source term for shielding design, particularly in lateral geometries.

The configuration without injection (Fig.~\ref{fig:MC_Sim_noAr}) shows that the overall radiation field is largely determined by the total electron emission rather than by the presence of a distinct high-energy bunch. Despite the negligible production of high-energy electrons in this case, the resulting dose distributions closely resemble those obtained with injection, consistent with the comparable total beam powers.

Overall, the configuration incorporating the full PIC phase-space distribution with injection reproduces the measured dose rates at the installed radiation monitors far more accurately than the initial idealized model. In particular, the electromagnetic dose component at the monitor position increases markedly when realistic large-angle and low-energy electrons are included (see Tab.~\ref{tab:tabledoses}). 
Quantitatively, the idealized simulation yields about \SI{60}{\micro\sievert\per\hour}, whereas the experimentally determined value during a \SI{13}{minute} long acceleration period is \SI{1097 \pm 73}{\micro\sievert\per\hour} (see Fig.~\ref{fig:dose_stability}). The full PIC-based simulation with injection (Case~3) results in \SI{1680}{\micro\sievert\per\hour}, thus reproducing the measured order of magnitude significantly better, particularly when accounting for beam power.

A direct comparison, however, is subject to several sources of uncertainty. The PIC simulation represents a single, idealized shot, while the experiment corresponds to a long-term average over many shots with varying laser and electron beam parameters. Furthermore, the charge of the main beam in the PIC simulation (\SI{24}{pC}, see Case~2) is higher than the value of about \SI{15}{pC} measured in the experiment \cite{ManuelKirchen2025FirstKALDERA}. In addition, the total emitted charge, and particularly the low-energy component emitted at large angles, cannot be measured directly and must be inferred from the PIC simulations, introducing further uncertainty. The Monte Carlo model is statistically robust, with uncertainties inside the tunnel on the order of \SIrange{1}{3}{\percent}, but is limited by its simplified geometry and interaction modeling.

Taken together, these effects limit the precision of a quantitative comparison. Nevertheless, the strong discrepancy between the idealized case and the experiment (60 vs.\ \SI{1100}{\micro\sievert\per\hour}) is substantially reduced when using the combined PIC and Monte Carlo approach. The remaining deviation between measurement and simulation is consistent with the expected uncertainties of the overall modeling chain.

\begin{table}[H]
\centering
\setlength{\tabcolsep}{12pt}      % more horizontal padding
\begin{tabular}{|l|l|r|r|r|}
\hline
\textbf{Case} & \textbf{Configuration} & \textbf{Beam Power} & \textbf{Charge} & \textbf{EM Dose} \\
\hline
Case 1 & Idealized & \SI{1.0}{W} & \SI{100}{pC} & \SI{60}{\micro\sievert\per\hour} \\
Case 2 & PIC (\SI{\pm 25}{mrad}) & \SI{0.26}{W} & \SI{24}{pC} & \SI{690}{\micro\sievert\per\hour} \\
Case 3 & PIC (full, inj.) & \SI{0.63}{W} & \SI{3}{nC} & \SI{1680}{\micro\sievert\per\hour} \\
Case 4 & PIC (full, no inj.) & \SI{0.60}{W} & \SI{4}{nC} & \SI{1210}{\micro\sievert\per\hour} \\
\hline
-- & Experiment & \SI{0.18}{W}$^{\dagger}$ & \SI{15}{pC}$^{\dagger}$ & \SI{1100}{\micro\sievert\per\hour} \\ \hline
\end{tabular}
\caption{Comparison of the FLUKA simulation configurations and the resulting electromagnetic dose component at the radiation monitor position inside the tunnel. The saturation-corrected average dose measured during commissioning experiments is included for comparison. \\
$^{\dagger}$ Experimental beam parameters \cite{ManuelKirchen2025FirstKALDERA} include only the measured on-axis electron component, as large-angle electrons were not accessible to diagnostics.}
\label{tab:tabledoses}
\end{table}

These results demonstrate that simplified assumptions based solely on a well-collimated, high-energy beam are insufficient for realistic radiation safety assessments in plasma accelerators. A detailed characterization of the complete electron emission, including low-energy electrons emitted at large angles, is therefore essential for a reliable and experimentally consistent description of the radiation environment.

Furthermore, the simulated dose levels of up to the order of \si{\gray\per\hour} in the immediate vicinity of the beamline can be related to the radiation damage thresholds discussed in Sec.~\ref{sec:raddamage}. While such dose rates remain localized and do not pose a limitation for standard operation, they indicate that non-radiation-hardened electronics in close proximity to the beamline may accumulate relevant doses over extended operation periods (hundreds to thousands of hours) if not adequately shielded. 

In addition, the simulations reveal a pronounced forward-directed electromagnetic shower originating from electron interactions with beamline components, which is capable of penetrating even dense shielding structures such as the concrete beam dump. This forward component is of particular relevance for the placement of sensitive devices downstream of the plasma stage, such as undulators with permanent magnets that are susceptible to radiation-induced demagnetization. 

In this context, the results highlight the importance of a targeted shielding concept and careful component placement, particularly in view of future upgrades towards higher repetition rates and average beam powers.

\section{Outlook and Future Work}
As plasma accelerators move toward higher repetition rates and higher average beam powers, radiation safety will become increasingly critical. In particular, the emission of electrons and associated radiation at large angles must be studied in more detail, both experimentally and through advanced simulations. Related efforts also exist in the context of beam-driven plasma accelerators, where plasma heating and its impact on radiation generation are being investigated \cite{I.Najmudin2025TemperatureAccelerator,J.Cowley2025DiagnosticsFlow}. These activities provide an important foundation and should be more closely connected to radiation protection studies for laser-driven plasma accelerators, in order to develop radiation protection concepts applicable to all types of plasma accelerators.

Future work should focus on a more detailed understanding of electron heating processes inside the plasma and their role in generating low-energy electrons emitted at large angles. Extended particle-in-cell simulations will be required to disentangle the underlying mechanisms and to establish robust correlations between plasma parameters, electron emission characteristics, and resulting radiation fields. At the same time, these studies should aim at developing reduced or semi-analytical source models that capture the essential features of large-angle electron emission. Such models would enable radiation transport simulations to be performed without relying exclusively on computationally expensive particle-in-cell calculations, thereby facilitating routine radiation safety assessments.

In addition, further investigations of beam dumping concepts for beam-driven plasma accelerators are required. The controlled disposal of the spent high-energy driver bunch remains a key technical challenge and is essential for enabling efficient shielding and limiting radiation damage to accelerator components. Dedicated beam dump and shielding designs tailored to the specific loss patterns of plasma accelerators will be crucial for future high-average-power facilities.

Overall, radiation protection must be integrated at an early stage into the design of future plasma accelerator facilities. Only by treating plasma dynamics, beam dumping, shielding, and component placement as tightly coupled design aspects can safe, reliable, and sustainable operation be ensured as plasma accelerator technology advances toward user-relevant applications.

\section{Summary}
Radiation safety in plasma accelerators shares many fundamental characteristics with conventional linear accelerators but is complicated by the distinctive electron beam parameters and by distributed beam losses, in particular from electrons emitted from the plasma at large angles. Bremsstrahlung and secondary neutron production remain the dominant radiation sources, with dose maxima occurring at electron energies well within the operating range of many plasma accelerator experiments. As a result, significant radiation fields can already arise at comparatively low electron energies, particularly in high-repetition-rate operation.

At KALDERA, the combination of dedicated radiation measurements with particle-in-cell and Monte Carlo simulations provides a comprehensive picture of the radiation fields generated during laser--plasma acceleration. This integrated approach enables identification of radiation contributions from electrons emitted from the plasma target at large angles with respect to the nominal beam axis. Particle-in-cell simulations reveal a substantial population of low-energy electrons distributed over wide angles, while Monte Carlo simulations show that their interactions with nearby structures contribute significantly to the measured radiation levels. Together, these results provide a consistent interpretation of both the magnitude and temporal structure of the observed dose rates and demonstrate that particle-in-cell--based source terms, coupled with Monte Carlo radiation transport, form a powerful framework for radiation studies at plasma accelerators.

As plasma accelerators progress toward higher repetition rates and average beam powers, radiation protection must be addressed as an integral part of facility design. In particular, radiation generation close to the plasma source, distributed beam losses, and the controlled disposal of accelerated beams require careful consideration. Addressing these aspects is crucial for enabling further increases in average beam power, which are a key prerequisite for the use of plasma accelerators in user-relevant and application-oriented settings. Future work should therefore focus on improved modeling of beam losses, the development of simplified radiation source models informed by plasma simulations, and the design of shielding and beam dump concepts specifically adapted to plasma-based accelerators.

%\section*{\label{sec:Ref}References}
%\bibliographystyle{iopart-num}
%\bibliography{references.bib}% Produces the bibliography via BibTeX.
%\printbibliography

\section*{Acknowledgements}
Generative AI tools (ChatGPT, OpenAI, GPT-5.x) were used to support the drafting and rephrasing of parts of this manuscript. All content was reviewed, edited, and validated by the authors, who take full responsibility for the accuracy and integrity of the work.

This research was supported in part through the Maxwell computational resources operated at Deutsches Elektronen-Synchrotron DESY, Hamburg, Germany
%\begin{comment}

%\end{comment}

\end{document}